\documentclass[11pt]{article}

\usepackage[margin=1in]{geometry}
\usepackage{times}
\usepackage{microtype}
\usepackage{graphicx}
\usepackage{booktabs}
\usepackage{amsmath, amssymb}
\usepackage{xcolor}
\usepackage{xspace}
\usepackage{hyperref}
\usepackage[section]{placeins}

\usepackage[numbers,sort&compress]{natbib}

\hypersetup{colorlinks=true, linkcolor=blue, citecolor=blue, urlcolor=blue}

\newcommand{\mlir}{MLIR\xspace}
\newcommand{\asyncdialect}{Async\xspace}
\newcommand{\dma}{DMA\xspace}
\newcommand{\tcm}{TCM\xspace}
\newcommand{\npu}{NPU\xspace}

\title{\bf Analyzing Latency Hiding and Parallelism in an MLIR-based AI Kernel Compiler}

\author{
  Javed Absar\textsuperscript{1},
  Samarth Narang\textsuperscript{2},
  and Muthu Baskaran\textsuperscript{2}\\
  \textsuperscript{1}Qualcomm Technologies International, Ltd.\\
  \textsuperscript{2}Qualcomm Technologies, Inc.\\
  \texttt{\{mabsar,samanara,muthub\}@qti.qualcomm.com}
}

\date{}

\begin{document}
\maketitle

\begin{abstract}
AI kernel compilation for edge devices depends on the compiler’s ability to exploit parallelism and hide memory latency
in the presence of hierarchical memory and explicit data movement, among other factors. This paper reports a benchmark
methodology and corresponding results for three compiler-controlled mechanisms in an \mlir-based compilation pipeline:
vectorization (Vec), multi-threading (MT) across hardware contexts, and double buffering (DB) using ping--pong scratchpad
buffers to overlap \dma transfers with compute. Using Triton/Inductor-generated kernels, we present an ablation ladder that
separates the contribution of Vec, MT, and DB, and we quantify how MT speedup scales with problem size using GELU as a
representative activation kernel. The results show that vectorization provides the primary gain for bandwidth-sensitive
kernels, MT delivers substantial improvements once scheduling overhead is amortized, and DB provides additional benefit when
transfers and compute can be overlapped (i.e., outside the extremes of purely memory-bound or purely compute-bound behavior).
\end{abstract}

\section{Introduction}
Kernel optimization is difficult to automate: hand-written kernels remain hard to beat, yet production systems require
portable and maintainable code generation across rapidly evolving architectures. On edge NPUs, performance is shaped by
hierarchical memory, explicit \dma-managed transfers, and the need to schedule work so compute stays busy while transfers
are in flight.

End-to-end model results are essential, but they often obscure \emph{which} mapping mechanisms improve performance and
\emph{why}. We therefore adopt a reproducible, kernel-oriented evaluation methodology aligned with Triton/Inductor-style
code generation, benchmarking representative operator kernels under controlled problem sizes to separately quantify the
impact of vectorization (Vec), multi-threading (MT), and double buffering (DB) \citep{tillet2019triton}. Vec exploits
data-level parallelism; MT exploits loop- and region-level parallelism by distributing independent tiles across hardware
contexts; and DB reduces stall time by overlapping memory transfers with compute.

To attribute gains to specific mechanisms, we report results using a simple ablation ladder:
\[
\texttt{Scalar} \rightarrow \texttt{Vec} \rightarrow \texttt{Vec+MT} \rightarrow \texttt{Vec+MT+DB}.
\]
In this ladder, \texttt{Vec} isolates SIMD-style lowering, \texttt{Vec+MT} quantifies incremental thread-level speedup,
and \texttt{Vec+MT+DB} evaluates whether an explicit latency-hiding schedule provides additional improvement once Vec and MT
are already in place. In addition to reporting results, we describe the concrete \mlir IR patterns and pass structure used
to realize MT and DB, making the methodology easy to reproduce and extend.

\section{Implementation Details: Multi-threading and Double Buffering}
\label{sec:impl}

Our implementation is built in \mlir \citep{lattner2020mlir} and follows a design principle: keep the intent expressed in
structured IR for as long as possible, and lower to runtime constructs only after the compiler has imposed a schedule.
This improves portability and makes transformations easier to validate.

\subsection{Multi-threading (MT)}
\label{sec:impl-mt}

\paragraph{Hardware model.}
The target \npu supports multi-threaded vector execution, where each hardware thread is associated with an independent
vector context (e.g., vector register and predicate state). This enables concurrent execution of independent tiles of the
same kernel, provided the tiled iteration space can be partitioned without cross-thread dependences.

\paragraph{Two-stage MT lowering.}
We implement MT as a two-stage pipeline that preserves structured parallelism and then introduces an explicit fork--join.
First, \emph{Form-Virtual-Threads} rewrites a tiled kernel (e.g., \texttt{linalg.generic}) into an explicitly parallel form
using \texttt{scf.forall}. The pass uses a size-based profitability heuristic over the tile space and selects a distribution
policy (block vs.\ block-cyclic) to balance work when ranges are uneven.

Second, \emph{Form-Async-Threads} lowers \texttt{scf.forall} to a fork--join representation using \mlir’s \asyncdialect
\citep{mlir_async_dialect}. Each tile becomes an \texttt{async.execute} region that produces a token; tokens are collected
into an async group, and \texttt{async.await\_all} forms a barrier before subsequent dependent computation.

\paragraph{Canonical fork--join skeleton in IR.}
The generated pattern is intentionally small and regular, which makes it straightforward to lower into a coroutine/task
runtime \citep{mlir_async_dialect}:

\begin{verbatim}
%group = async.create_group %N
scf.for %tile = ... {
  %tok = async.execute {  /* tile body */  async.yield }
  async.add_to_group %tok, %group
}
async.await_all %group
\end{verbatim}

\paragraph{Why \asyncdialect (instead of lowering MT directly).}
Keeping MT as a structured fork--join in \asyncdialect preserves parallel semantics in a declarative form until late
lowering, while still enabling a straightforward translation to runtime scheduling \citep{mlir_async_dialect}.

\subsection{Double buffering (DB)}
\label{sec:impl-db}

Double buffering is a software-pipelining strategy that overlaps transfers and compute by alternating between two
scratchpad buffers (ping and pong). The technique is closely related to modulo scheduling and pipelined loop execution
\citep{lam1988sps,rau1994iterative}. We implement DB in two stages, separating the construction of a pipelined schedule
from the introduction of target-specific asynchronous transport primitives.

\paragraph{Stage 1: structural pipelining.}
Stage~1 matches a single-buffered tiled-loop “normal form” typically created by tiling:
\[
\texttt{memref.subview} \rightarrow \texttt{memref.alloc} \rightarrow \texttt{memref.copy},
\]
followed by compute and then a write-back sequence. From this structure, the pass builds an explicit ping--pong schedule.
It emits a prologue that prefetches the first tile into ping buffers, then rebuilds the main loop into two alternating
sub-kernels. Each sub-kernel (i) issues a prefetch of the \emph{next} tile into the opposite buffer, (ii) computes using
the current buffer, and (iii) reconstructs storeback by rematerializing subviews at the current induction variable. A
boolean toggle selects ping vs.\ pong each iteration, and the pass attaches lightweight attributes as anchors so the next
stage can reliably identify prefetch/compute/storeback regions.

\paragraph{Stage 2: asynchronous DMA integration.}
Stage~2 replaces synchronous copies with explicit asynchronous \dma operations. Prefetch paths are rewritten to
\texttt{memref.dma\_start} with distinct ping/pong tags, and \texttt{memref.dma\_wait} is inserted immediately before compute
to ensure tile residency in \tcm. Storeback copies can be handled similarly, and the pass emits balanced tag deallocations.
By staging the transformation, we first establish a correct schedule and only then map it onto the target’s transport
interface, mirroring classic compiler practice for pipelined loops \citep{lam1988sps,rau1994iterative}.

\paragraph{Composing DB with MT.}
DB composes naturally with MT. Once \texttt{memref.dma\_wait} enforces that a tile is resident in \tcm, the compute region
can exploit \texttt{scf.forall} (virtual threads) and/or the lowered fork--join representation to execute multiple sub-tiles
concurrently. This is the composition used to realize the \texttt{Vec+MT+DB} rung of our ladder.

\section{Benchmark Setup and Methodology}
We evaluate two representative kernels. The first is a bandwidth-centric 2D vector addition microbenchmark with shape
$[64, 128 \times 128]$ elements, which is useful for understanding the relative impact of vectorization and latency-hiding
in a memory-heavy regime. The second is GELU, a common activation kernel representative of transformer inference subgraphs;
we use a Triton implementation of GELU as our concrete kernel instance.

For the vector-add microbenchmark, we evaluate the ladder variants \texttt{Scalar}, \texttt{Vec}, \texttt{Vec+MT}, and
\texttt{Vec+MT+DB}. For GELU, we report end-to-end latency across a problem-size sweep for both single-threaded and
multi-threaded execution to expose MT overhead amortization and scaling. We report latency in microseconds ($\mu$s) and,
where applicable, speedup as the ratio of single-thread to multi-thread time.

\section{Results}
\subsection{vec-add-2d Ablation Ladder}
\begin{figure}[!htbp]
  \centering
  \includegraphics[width=\linewidth]{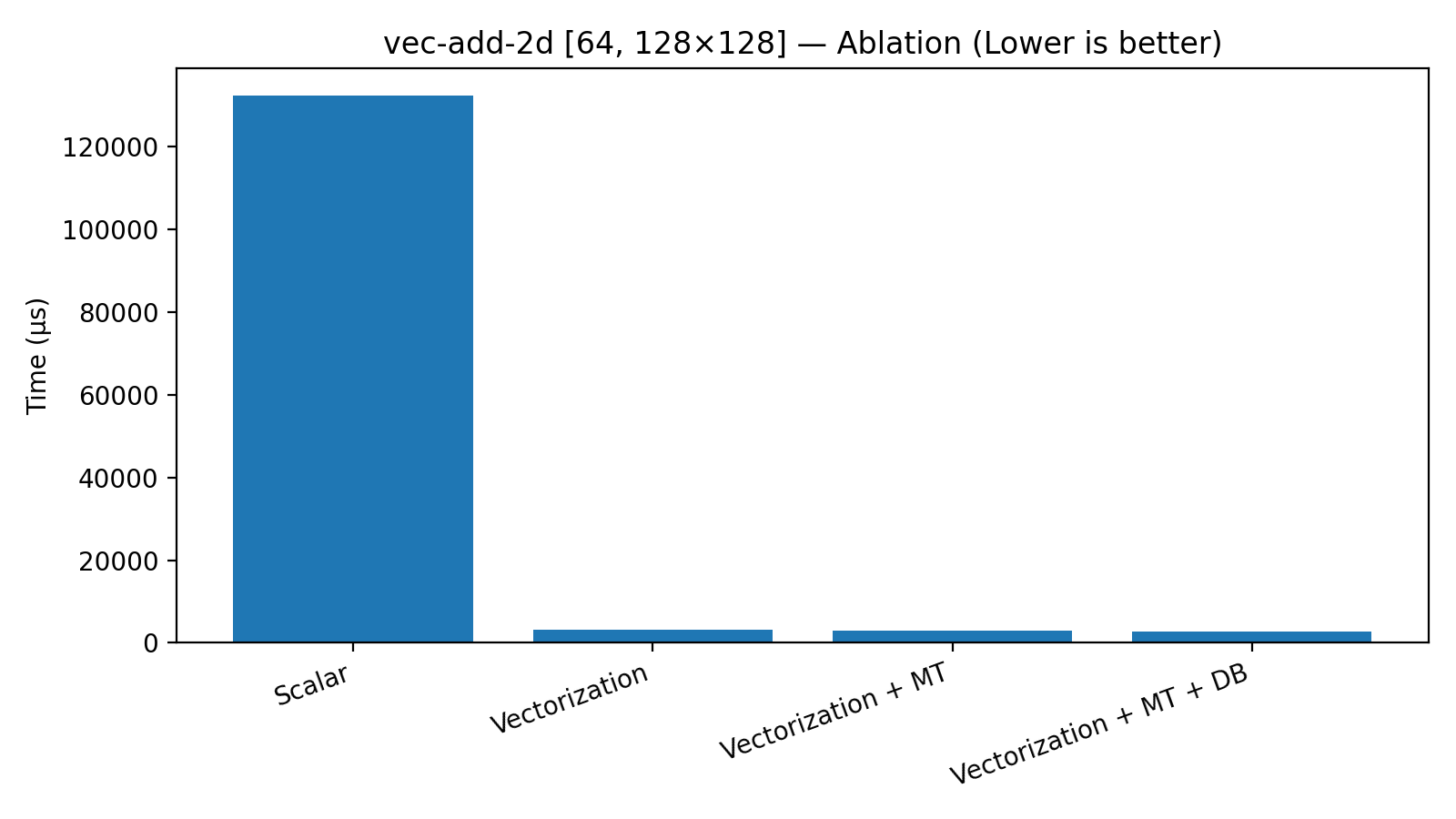}
  \caption{vec-add-2d ($[64, 128 \times 128]$ elements) ablation ladder:
  Scalar (132{,}479 $\mu$s), Vec (3{,}210 $\mu$s), Vec+MT (3{,}000 $\mu$s), Vec+MT+DB (2{,}689 $\mu$s).}
  \label{fig:vecadd}
\end{figure}

Figure~\ref{fig:vecadd} summarizes the ladder for vec-add-2d. Vectorization accounts for the dominant improvement
(132{,}479 $\mu$s $\rightarrow$ 3{,}210 $\mu$s, $\sim$41.3$\times$), consistent with a bandwidth-oriented kernel that benefits
immediately from SIMD-style lowering. MT and DB provide smaller but measurable incremental gains
(3{,}210 $\mu$s $\rightarrow$ 3{,}000 $\mu$s $\rightarrow$ 2{,}689 $\mu$s), suggesting that once vectorization is in place,
remaining headroom comes from reducing synchronization overheads and partially overlapping transfer/compute effects rather
than from increasing arithmetic throughput.

\subsection{GELU: Single-thread vs Multi-thread Scaling}
\begin{figure}[!htbp]
  \centering
  \includegraphics[width=\linewidth]{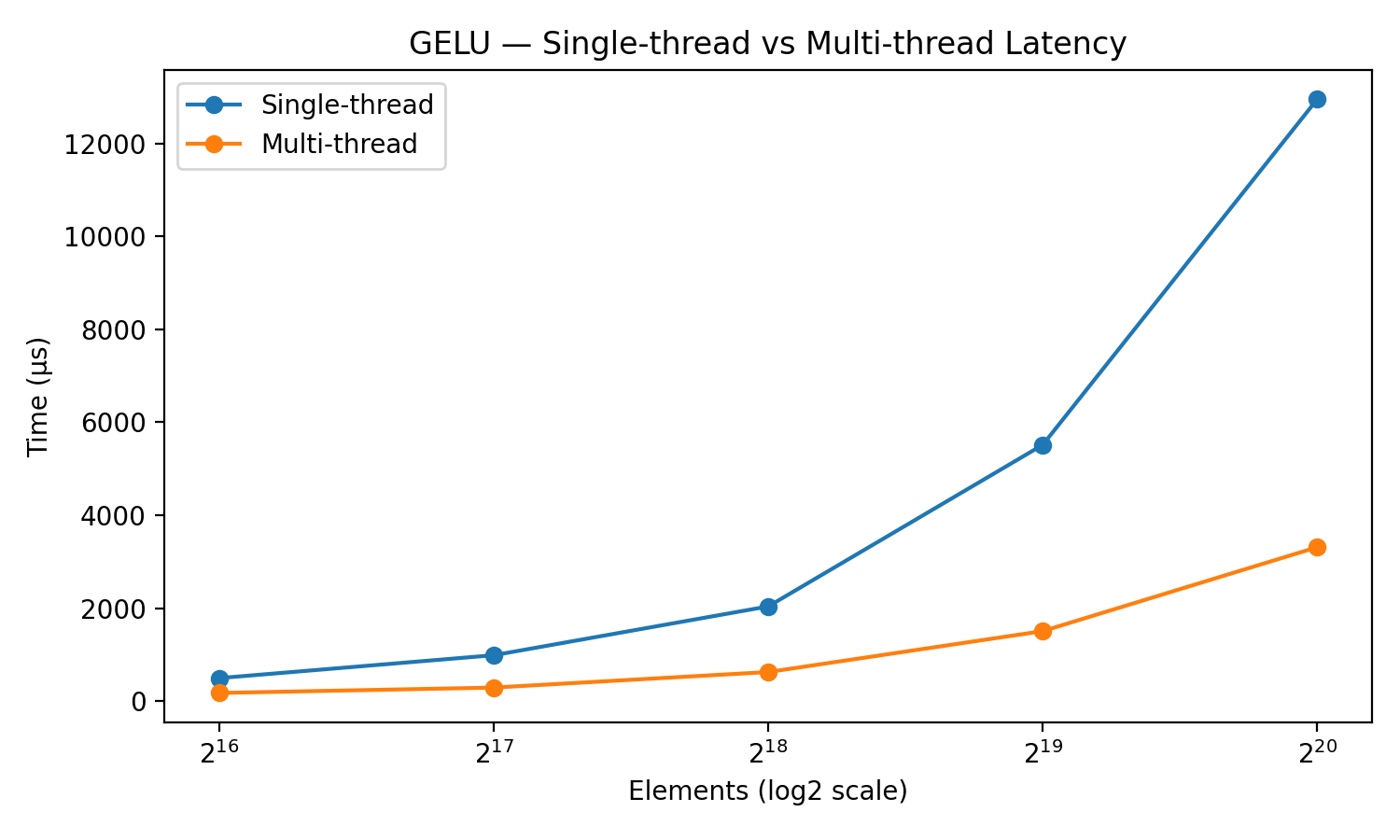}
  \caption{GELU latency vs problem size for single-threaded and multi-threaded execution.}
  \label{fig:gelu_latency}
\end{figure}

\begin{figure}[!htbp]
  \centering
  \includegraphics[width=\linewidth]{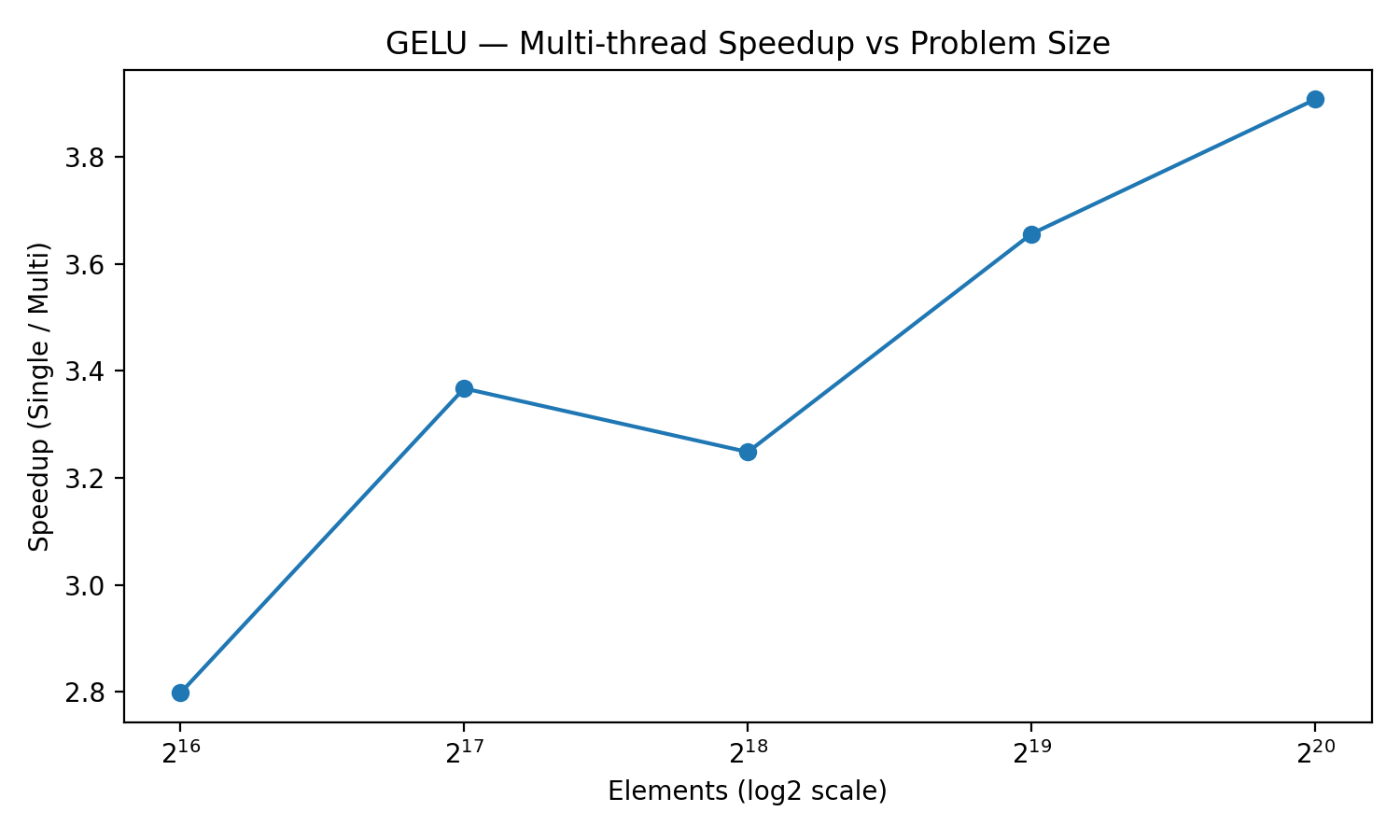}
  \caption{GELU multi-thread speedup (Single/Multi) vs problem size.}
  \label{fig:gelu_speedup}
\end{figure}

Figure~\ref{fig:gelu_latency} reports GELU latency over a size sweep and Figure~\ref{fig:gelu_speedup} shows the
corresponding multi-thread speedup. MT improves performance across all tested sizes and the speedup grows with problem size,
reaching $\sim$3.91$\times$ at 1{,}048{,}576 elements (12{,}947 $\mu$s single-thread vs 3{,}313 $\mu$s multi-thread). This
trend indicates that fixed overheads associated with fork--join scheduling are amortized as per-thread work increases, while
saturation at larger sizes suggests emerging shared bottlenecks such as memory bandwidth or barrier costs.

\subsection{Discussion}
Across these kernels, vectorization provides the key first-order win, especially in bandwidth-sensitive regimes. MT then
yields substantial additional improvement when there is sufficient parallel slack and the kernel is large enough to amortize
scheduling overhead. DB provides incremental benefit when transfers and compute are both significant; in the purely memory-
bound limit, DB is constrained by transfer bandwidth, and in the purely compute-bound limit, overlap opportunities are
limited.

\section{Conclusion}
We presented a compact, reproducible methodology for analyzing vectorization, multi-threading, and double buffering in an
\mlir-based kernel compiler. The proposed ladder makes it easier to attribute performance improvements to specific compiler
mechanisms rather than reporting only end-to-end speedups. Our measurements show that vectorization is foundational, MT can
provide large gains once overhead is amortized, and DB adds incremental benefit when transfer/compute overlap exists. Future
work will broaden the benchmark set (e.g., RMSNorm, softmax) and connect these measurements to a predictive model that
relates overlap efficiency to \dma throughput and scratchpad capacity.

\FloatBarrier
\bibliographystyle{unsrtnat}
\bibliography{refs}

\end{document}